\documentclass[letterpaper]{article}
\usepackage{aaai25}
\usepackage{times}
\usepackage{helvet}
\usepackage{courier}
\usepackage[hyphens]{url}
\usepackage{graphicx}
\usepackage{natbib}
\usepackage{caption}
\usepackage{algorithm}
\usepackage{algorithmic}
\usepackage{newfloat}
\usepackage{listings}

\graphicspath{ {./images/} }
\usepackage{csquotes}

\DeclareCaptionStyle{ruled}{labelfont=normalfont,labelsep=colon,strut=off}
\floatstyle{ruled}
\newfloat{listing}{tb}{lst}{}
\floatname{listing}{Listing}

\title{Toward Needs-Conscious Design: Co-Designing a Human-Centered Framework for AI-Mediated Communication}

\author {
    Robert Wolfe\textsuperscript{\rm 1}\thanks{Work performed while at the University of Washington},
    Aayushi Dangol\textsuperscript{\rm 2},
    JaeWon Kim\textsuperscript{\rm 2},
    Alexis Hiniker\textsuperscript{\rm 2}
}
\affiliations {
    \textsuperscript{\rm 1}Rutgers University\\
    \textsuperscript{\rm 2}University of Washington\\
    robert.wolfe[at]rutgers.edu, adango[at]uw.edu, jaewonk[at]uw.edu, alexisr[at]uw.edu
}

\begin{document}

\maketitle

\begin{abstract}
  We introduce \textit{Needs-Conscious Design}, a human-centered framework for AI-mediated communication that builds on the principles of Nonviolent Communication (NVC). We conducted an interview study with $N$=14 certified NVC trainers and a diary study and co-design with $N$=13 lay users of online communication technologies to understand how NVC might inform design that centers human relationships. We define three pillars of Needs-Conscious Design: Intentionality, Presence, and Receptiveness to Needs. Drawing on participant co-designs, we provide design concepts and illustrative examples for each of these pillars. We further describe a problematic emergent property of AI-mediated communication identified by participants, which we call \textit{Empathy Fog}, and which is characterized by uncertainty over how much empathy, attention, and effort a user has actually invested via an AI-facilitated online interaction. Finally, because even well-intentioned designs may alter user behavior and process emotional data, we provide guiding questions for \textit{consentful} Needs-Conscious Design, applying an affirmative consent framework used in social media contexts. Needs-Conscious Design offers a foundation for leveraging AI to facilitate human connection, rather than replacing or obscuring it.
\end{abstract}
\section{Introduction}

In April 2024, Meta deployed a chatbot backed by the LLaMA-3 language model across its portfolio of apps, including Facebook, Instagram, WhatsApp, and Messenger \cite{metaai}. Just as the first cohort of generative chatbots now mediates access to \textit{information} for many users via ChatGPT \cite{openai2022chatgpt} and the language models integrated into Bing and Google Search \cite{memon2024search}, it's clear that the second wave will mediate person-to-person \textit{communication}. Yet social media, a technology ostensibly designed to foster interpersonal relationships \cite{ten2017effect}, appears to contribute not to a sense of connection but to a sense of isolation for many users \cite{primack2019positive}, with the U.S. surgeon general warning of an ``epidemic of loneliness and isolation'' \cite{surgeongeneral2023}. For AI to have a prosocial effect - and avoid amplifying existing problems - will require intentional design that centers human interpersonal needs, rather than offering a technological substitute for connection.

In this research, we turn to a well-established model for communication that centers human needs: that of Nonviolent Communication \cite{rosenberg2003life}, or NVC, a process-oriented approach to empathetic communication positing that human conflict and feelings of isolation arise from unmet needs, and that communicating those needs without judgment can  yield deeper connections between people \cite{rosenberg2015nonviolent}. NVC originated in Rogerian person-centered psychology \cite{rogers1963concept} and now sees wide application in conflict mediation, much of which is overseen by the Center for Nonviolent Communication (CNVC), which certifies trainers who carry out workshops teaching the fundamentals of NVC. We build on NVC to co-design a framework to designing for human interpersonal needs that we call Needs-Conscious Design. We address three research questions:

\begin{enumerate}
    \item \textbf{RQ1}: How do NVC trainers envision NVC could be leveraged in AI that supports empathetic connection?
    \item \textbf{RQ2}: What designs do lay users envision to support empathetic connection, a core construct of NVC?
    \item \textbf{RQ3}: Building on RQ1\&2, what are the pillars of Needs-Conscious Design, and how can they be implemented?
\end{enumerate}

\begin{figure*}
\centering
\includegraphics[width=.68\textwidth]{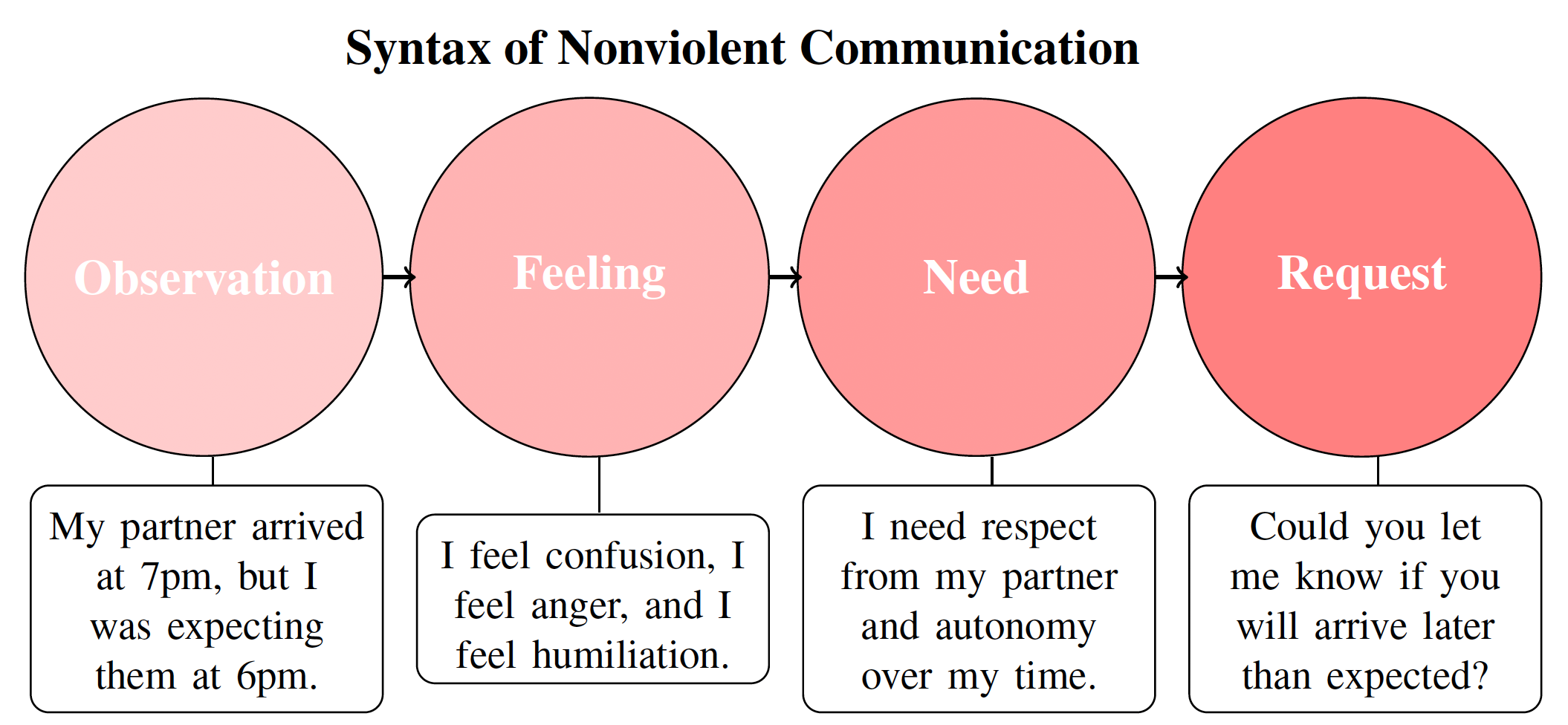}
\caption{An illustration of the ``syntax'' of NVC, with an example written by the authors. NVC emphasizes observing without judgment, becoming aware of feelings, tracing feelings to underlying needs, and requesting that those needs be satisfied.}
\label{fig:nvc_process}
\end{figure*}

\noindent To answer these questions, we enrolled $N$=14 trainers certified by the CNVC in an interview study. In parallel, we enrolled $N$=13 lay users of online communication technologies in a six-day diary study bookended with an entry interview at the beginning and an exit survey and co-design at the end. We intended to capture not just high-level design perspectives but also the everyday experiences of when users seek empathy online. We applied a deductive-inductive coding approach \cite{azungah2018qualitative} to the interviews with CNVC trainers and the responses and designs of lay users to derive a set of themes \cite{braun2012thematic} to answer our research questions. We make the following contributions:

\begin{enumerate}
    \item \textbf{We offer a conceptual model for Needs-Conscious Design, defined by the three pillars of Intentionality, Presence, and Receptiveness to Needs}. We provide core design considerations drawn from these three pillars for designers seeking to facilitate empathetic connection, leveraging the co-designs of lay users to provide examples.
    \item \textbf{We define Empathy Fog, an emergent property of AI-mediated communication}, wherein using generative AI to send a response obscures the effort and attention offered by a sender to a recipient. We find that empathy fog can also render the sender themselves uncertain of what they have invested in a relationship.
    \item \textbf{We offer guiding questions for the practice of consentful Needs-Conscious Design}, drawing on the affirmative consent framework of \citet{im2021yes}. We contend that design involving fluent, humanlike language technologies and user emotional data must center user consent.
\end{enumerate}

\noindent Needs-Conscious Design offers an approach that values human-to-human connection and the satisfaction of needs, providing a foundation for self-reflective designs that leverage AI not as a replacement for human-to-human connection, but as a facilitator and supporter of such connection.
\section{Related Work}\label{sec:rw}

\subsection{Nonviolent Communication}

Nonviolent Communication, or NVC, is an approach to structured communication intended to meet human needs \cite{rosenberg2015nonviolent}. As shown in Figure \ref{fig:nvc_process}, NVC employs a four-component approach to communication, characterized by 1) observing the situation without making judgments; 2) accurately naming one's own feelings; 3) linking feelings to the underlying needs to which they are tied; and 4) making requests for those needs to be met \cite{rosenberg2015nonviolent}. While NVC grew out of Rogerian person-centered psychology \cite{rogers1963concept,rogers1963toward}, it has seen much application in the realm of conflict mediation, as NVC was implemented to facilitate racial integration of American schools \cite{rosenberg2015nonviolent}, and it has been used, sometimes alongside mindfulness training \cite{suarez2014freedom}, in schools \cite{rosenberg2003life}, in prisons \cite{marlow2012nonviolent, suarez2014freedom}, and in healthcare settings \cite{museux2016improving, nosek2014nonviolent, wacker2018preventing}.

\subsection{AI-Mediated Communication}

Recent work suggests that AI-driven language technologies, and especially chatbots like ChatGPT \cite{openai2022chatgpt}, can help to \textit{mediate} human communication by acting ``on behalf of a communicator by modifying, augmenting, or generating messages to accomplish communication goals'' \cite{Hancock2020AIMediatedCD}. Research positioning AI as a mediator of human relationships includes that of \citet{Zhang2023IceBreakingTR}, who find that AI can provide ``conversation facilitators'' that break the ice and engender deep conversations between strangers. Similarly, \citet{Shin2023IntroBotET} find that a chatbot can facilitate online discussions by using the social media data of users to familiarize users with each other. \citet{fu2023text} find users prefer AI-mediated communication tools more in formal than in informal situations. 

AI-Mediated Communication can have undesirable effects on trust: \citet{Jakesch2019AIMediatedCH} show that users mistrust AirBnB profiles perceived as AI-written when examining AI and human profiles, while \citet{Liu2022WillAC} find that trust decreases among recipients of emails who were told that AI was involved in the writing process. Moreover, \citet{Hohenstein2020AIAA} find that AI is often blamed when AI-mediated conversations go awry, displacing to AI some of the responsibility attributable to human communicators.

In addition to research positioning AI as a mediator, \citet{Drrenbcher2023TheIO} contend that much research conceptualizes AI as a \textit{substitute} for human relationships. For example, \citet{Xygkou2023TheA} describe seven scenarios in which chatbots might be employed to offer emotional support to mourners during a time of grief, including acting as a friend, listener, emotion coach, romantic partner, or simulation of the deceased. Increasing in human-like presentation, \citet{Dagan2021SynergisticST} design a system with its own ``needs'' for social interaction. In some cases, AI may be used to stand in for a specific person. \citet{Lee2023SpeculatingOR} find that such ``AI clones'' of a person can result in negative reaction to displaced identity, threats to self-perception and individuality, and over-attachment to the AI clone of someone that one already knows.

\subsection{Design Frameworks for Human Connection}

That our research introduces a framework for designing to support human connection places it in conversation with prior work introducing frameworks for similar human-centered purposes. \citet{baughan2021someone} introduce Interpersonal Design, an approach that centers relationships in the design of technologies, in the context of designing for conflict resolution and good faith disagreement. Similarly, \citet{zhang2021designing} propose designing for emotional well-being, an approach that posits ways to nudge users into prosocial interactions likely to increase their well-being, so long as those interactions do not feel emotionally burdensome, especially for users struggling with mental health. \citet{Marcu2023AttachmentInformedDD} additionally introduce Attachment-Informed Design, a set of principles for supporting both relationships and communities in the context of mental health interventions. Finally, \citet{kelly2017demanding} introduce Effortful Communication, suggesting that designing for interpersonal relationships should support communication that is ``effortful'' or ``demanding by design'' and characterized by ``discretionary investment, personal craft, focused time, responsiveness to the recipient, and challenge to a sender's capacities,'' which is often deeply meaningful to the recipient of the communication. In a study of an iOS app allowing partners to invest effort into digital experiences, \citet{zhang2022auggie} found that \textit{personally} effortful communication is distinct from \textit{procedurally} effortful communication arising from barriers to using a system; that is, a design can support communication evidencing effort and craft, but remain convenient and usable to the user. 

\subsection{Approaches to Emotional Data}

AI-Mediated Communication comes with the risk that a user's interpersonal data will be collected and used for purposes not intended by the user. This is a particular concern in the context of \textit{empathetic} communication, which may elicit emotional data that could be used in deceptive or compromising ways \cite{gray2018dark,mathur2021makes}. \citet{solove2023data} contend that regulation should expand from categorizing types of personal data to consider the \textit{harms} that could arise from misuse of such data. 

Recent research seeks to prompt designers of applications to consider how their systems can provide users with both agency and the ability to meaningfully consent. \citet{lukoff2021design} propose design that centers user sense of \textit{agency}, demonstrating how this might be implemented on video sharing platforms like YouTube. Similarly, \citet{petridis2022tastepaths} introduce a graph-based recommender system that provides greater exposure to the decisions of the system to the user, enabling the user to provide greater feedback and recognize why recommendations were made. \citet{im2021yes} propose the application of affirmative consent as a framework for addressing problems with social platforms, including those related to algorithmic content curation;  we draw on this framework to pose guiding questions for consentful Needs-Conscious Design. Finally, \citet{chowdhary2023can} question whether users can meaningfully consent to wellbeing technologies implemented in the workplace.
\section{Methods}

We conducted a diary study with $N$=13 participants who reported using text-based online communication at least daily, and an interview study with $N$=14 CNVC-certified trainers. Our University's IRB approved this research. 

\subsection{Diary Study}

We conducted a four-phase diary study with $N$=13 participants, notated with the prefix \textit{P} (\textit{e.g.,} ``P3 said\dots''). We ran a pilot with a member of the research team and with a relative of a team member, allowing us to correct issues with data collection forms and revise confusing interview questions.

\subsubsection{Entry Interview}

We designed a ten-question structured interview protocol. Four questions asked about a time participants had 1) extended empathy in text online; 2) received empathy in text online; 3) sought but not received empathy in text online; and 4) not extended empathy to someone who sought it in text online. Questions sought to elicit the interpersonal needs participants hoped to satisfy online, and scenarios in which they did not feel comfortable interacting online. Remaining questions asked about the relative difficulty of empathetic connection online vs. in person; easy vs. difficult social situations to connect with empathy online; whether some platforms make it easier to extend empathy; times when showing empathy is less important online; and what communication styles encourage empathy online.

\subsubsection{Diary Study}

After the interview, participants were asked to submit two diary entries per day via a Google Form. The first entry would describe a time during the day when the participant sought empathy online, and the second a time when someone sought empathy from the participant online. Participants summarized these interactions and shared the full text of conversations (with names and PII redacted) if they were comfortable. We gathered 109 total entries, an average of 8.4 per participant (about 1.4 per day).

\subsubsection{Exit Survey}

The exit survey asked participants about their experiences during the the diary study. Participants were asked to describe when they sought empathy and received it, and when they did not receive it; when it was easy to think in terms of needs and feelings underlying online interactions, and when it was not easy; with whom it was easiest and hardest to converse with empathy; when they chose not to communicate their needs and feelings; aspects of the online environment that made it easier or harder to communicate empathetically; and what stood out most from the week. Questions mirrored the entry interview to facilitate observations only available after six days of contextual inquiry \cite{raven1996using}. A second page primed participants for co-design by asking them to reflect on thirteen technologies envisioned by the study team based on NVC teaching methods described by \citet{rosenberg2015nonviolent}.

\subsubsection{Co-Design}

The study concluded with a co-design session via Zoom video call. Participants met individually with the first or second author and reflected aloud on design features to facilitate more empathetic communication online. After arriving a design, the participant sketched a low-fidelity prototype using pen and paper or Zoom Whiteboarding. The participant then reflected on features to facilitate \textit{less} empathetic communication, to elicit designs to \textit{avoid}. Participants then sketched a low-fidelity prototype of this technology.

\subsubsection{Participants}

We posted study advertisements to Reddit, LinkedIn, Facebook, and Slack channels at our University. We enrolled 15 participants who met our inclusion criteria. Upon enrollment, all participants were provided with a consent form outlining the study procedures and providing detailed information about payment and study timeline. \looseness=-1

\begin{table}[htbp]
    \centering
    \small
    \begin{tabular}{|c|c|c|}
        \hline
         Study Phase & Time Commitment & Compensation \\
         \hline
         Entry Interview & 30-40m & \$10 \\
        Diary Entries & 10-20m/day & \$5/day (\$30) \\
         Exit Survey & 30-40m & \$10 \\
         Co-Design Session & 30-40m & \$25 \\
         \hline
    \end{tabular}
    \caption{The compensation schedule used for the diary study.}
    \label{tab:participant_compensation}
\end{table}

\noindent One participant completed only the entry interview, and a second completed the entry interview and one day of the diary study. We compensated these participants for the parts of the study they completed. $N$=13 participants completed all phases of the study. We used pilots to estimate participant time commitment and adjusted compensation according to the schedule in Table \ref{tab:participant_compensation}. Compensation for the co-design session was higher to motivate participants to complete the full study. Participants received Amazon gift credit, not cash, and could earn an additional \$5 by explaining their reasoning in the exit survey. Table \ref{tab:participant_demographics} describes the demographics of our study participants. We happened to over-sample people who identify as women and people aged 25-34.

\begin{table}[htbp]
\small
    \centering
    \begin{tabular}{|c|p{6cm}|}
        \hline
        Category & Participant Demographics ($N$=13) \\
         \hline
         Gender & 7 Women, 3 Men, 1 Nonbinary, 1 Woman and Nonbinary, 1 Prefer Not to Say \\
         Race & 5 Asian, 3 Black, 3 White, 1 Asian and White, 1 Prefer Not to Say \\
         Ethnicity & 12 Not Hispanic or Latino, 1 Prefer Not to Say \\
         Age Range & 7 25-34, 3 35-44, 1 18-24, 1 45-54, 1 Prefer Not to Say \\
         \hline
    \end{tabular}
    \caption{Self-reported demographics of diary participants.}
    \label{tab:participant_demographics}
\end{table}

\subsection{CNVC Trainer Interview Study}

We conducted an interview study with $N$=14 CNVC-certified trainers, notated with the prefix \textit{T} (\textit{e.g.,} ``T3 said\dots''). Interviews lasted between 30 minutes and two hours, with most interviews lasting approximately one hour.

\subsubsection{Interview Protocol}

We created a semi-structured interview protocol that asked trainers about the following:

\begin{enumerate}
    \item The NVC model (Observation, Feeling, Need, Request).
    \item Strategies and exercises for effectively using NVC.
    \item What successful NVC looks like, and how trainers understand success in trainees.
    \item Challenges in learning and effectively using NVC.
    \item How NVC might inform online communication, and limitations of online settings for employing NVC.
    \item What designs for more empathetic online communication might arise from NVC.
\end{enumerate}

\noindent We also asked clarifying questions when participants raised other topics germane to the study (\textit{e.g.}, when T14 suggested using generative AI in NVC training).

\subsubsection{Participants}

We used the CNVC website to filter a list of certified NVC trainers in the U.S. who spoke English, in line with our IRB. We then reached out to 30 trainers via email, explaining our interest and providing a study overview. We provided these expert participants \$50 Amazon credit and did not request that they provide demographic information.

\subsection{Data Analysis}

The study team applied a deductive-inductive approach \cite{azungah2018qualitative} to coding the study data using Atlas.ti qualitative analysis software \cite{atlas2024}. Deductive codes mirrored our research questions and included: Needs supported online (RQ1), Self-care and Self-empathy (RQ1), Patterns of empathetic communication (RQ2), Patterns of non-empathetic communication (RQ2), Challenges in empathetic communication online (RQ2), and Design Considerations (RQ3). We generated inductive subcodes within these deductive codes (\textit{e.g.}, ``Needing a Space to Vent'' was a subcode for the ``Needs supported online'' deductive code). \looseness=-1

The first two authors and the last author together coded all data collected from two diary study participants and met to discuss the initial set of inductive subcodes. The first two authors then independently coded data submitted by one diary study participant each. They then met to propose new subcodes, merge redundant subcodes, and exchange notes and relevant quotes. They repeated this process five times to code all participant data, before meeting with the last author to discuss the final codebook. The same process was applied to code the CNVC trainer interviews, with data from two participants first coded jointly, and the remaining twelve coded by the first two authors, two participants at a time. Using the deductive and inductive codes, the first two authors followed a thematic analysis process \cite{braun2022conceptual} and wrote memos on themes answering the research questions. They met to discuss the themes, grouping them according to question, with explanatory notes and supporting quotes. Finally, the first two authors and the last author met and agreed on the set of final themes reflected in the Findings.
\section{Findings}\label{sec:findings}
Participants surfaced three common design principles: \textit{intentionality}, \textit{presence}, and \textit{receptiveness to needs}. We describe each of these attributes and the ways participants envisioned technology fostering or impeding each.

\begin{figure*}
\centering
\includegraphics[width=.8\textwidth]{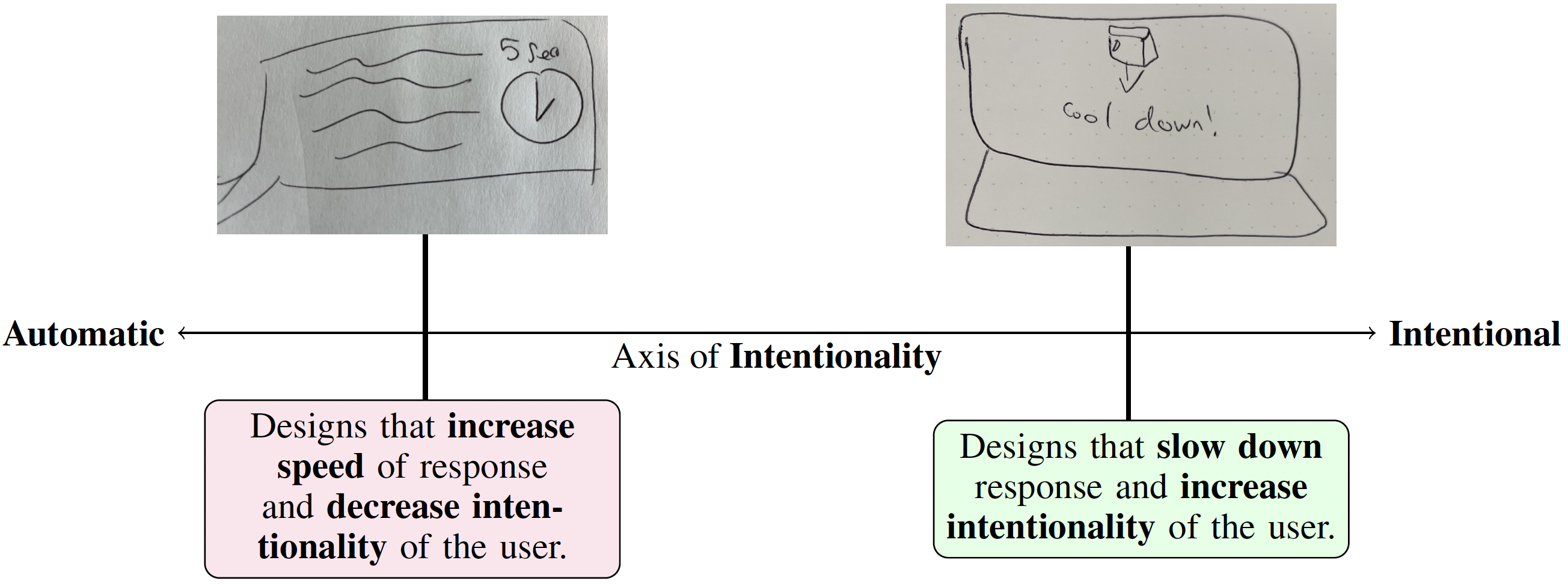}
\caption{Needs-Conscious Design requires that users feel they have communicated intentionally. Participants preferred design that supports \textbf{slowing down} for self-reflection and to improve sense of agency. \textbf{Top right}: Co-design from P10, an animated ice cube melting, reminding the user to slow down before responding in a way they might regret. \textbf{Top left}: Co-design from P12, a timer ticking down to a message being deleted if no response is sent, increasing automaticity and decreasing self-reflection.}
\label{fig:intentionality_axis}
\end{figure*}

\subsection{Pillar 1: Designing for Intentionality}

Both trainers and diary study participants identified \textit{Intentionality} as central to forming empathetic connections with others online. Trainers explained that practicing NVC entails much more than following a script; it requires intentionally choosing to prioritize needs, communicating slowly enough to be intentional about the word choice, and intentionally valuing one's own needs (as well as the needs of others).

\subsubsection{Consciously Choosing NVC} Trainers stressed that NVC requires an intentional commitment to centering \textit{needs}, both of self and others. NVC provides a structure for communication, but trainers stressed that following this structure is insufficient without \textit{Nonviolent Consciousness} or \textit{Needs Consciousness}. Every trainer highlighted the difference between following the structured format of NVC (\textit{i.e.}, the Observations, Feelings, Needs, Requests [OFNR] formula) and Needs Consciousness. T12 explained that the structure of NVC is just \textit{``a framework that helps you get to\ldots NVC consciousness, which is focusing on, `I care more about my connection with you, and also my with myself.'''} Trainers said that without Needs Consciouness, the NVC structure can be abused. T14 said that \textit{``if you miss the consciousness part, then my intention could be to get my way, or my intention could be to manipulate a certain outcome. And then I'm using words that sound like NVC, but now I've weaponized NVC and it's not NVC at all. So the consciousness and the intentionality is really primary.''} Similarly, T3 said, \textit{``NVC can be a tool that can contribute to harm if not used with the mindset of seeking connection\ldots it's very important to use NVC with a lot of intentionality.''} T7 said, \textit{``that's a core value of NVC response, authenticity, empathy, responsibility, shared power, and choice\ldots not to get my agenda over on you, get your buy-in, or coerce you into my agenda.''}

\subsubsection{Slowing Down} Trainers and diary study participants alike explained rapid-fire communication impeded intentionality, while slower, asynchronous back-and-forth facilitated self-reflection. T14 explained, \textit{``The faster the medium, the less conducive it is to connection.''} Several diary study participants described slowing down to become more mindful of their feelings and needs. During the entry interview, P1 described the importance of \textit{``checking in''} with oneself when emotions run high: \textit{``I think when it's distressing, your fight-or-flight response is kicking in, and I mean, you can't run away from the thing that's in your pocket.''} During the co-design, P10 reflected on the inflammatory effects of platform-induced speed in online spaces, saying, \textit{``I think that's one of the biggest contributors to unempathetic conflict online. It's just the speed of it all\ldots when I've had unempathetic conversations with family members in the past, the speed of it, and the algorithm was designed to then show a public conversation to other people with their like-minded view\ldots It just felt like a wrestling match with everybody in the ring.''} Without algorithmic incentives motivating a fast response, participants said text-based communication could help slow down and respond with intention. P9 said, \textit{``I can choose to react in my own way, and in my own time, and choose my words a little more carefully.''} P2 said that \textit{``when your emotions are running high or you're just lashing back as a reaction, it's harder to stop that in person than with online communication. You can really take more time to step away or rethink what your words are before you send them.''}

\subsubsection{Self-Empathy}

Both trainers and participants described the relationship to oneself as an essential part of having an intentional, empathetic relationship with another person online. T6 identified that the first step in learning NVC as \textit{``an internal application of Nonviolent Communication, sometimes known as self-empathy or self-connection.''} P3 described strategies like taking walks to practice self-reflection: \textit{``Don't send this message. Calm down. Get a walk. Yeah, go out for five minutes, and then come back, and maybe think [of] a better way of getting this message across.''} In the exit survey, P7 expressed the difficulty of extending empathy \textit{``when you already have emotional distress\dots it is difficult to communicate with empathy.''}

Other participants noted the importance of safety and reciprocity. P3 explained that extending empathy to someone who has \textit{``an agenda''} can backfire and harm the person who is trying to be empathetic. P6 noted they set boundaries when requests involve \textit{``unreasonable resources from the outside, like lending money, or like, coming to your house for a night.''} During the entry interview, P1 noted the importance of caring for oneself when a boundary is crossed:\looseness=-1

\begin{displayquote}
\textit{``If somebody crosses a boundary, then you need to extend that kindness to yourself. You don't stop having empathy for the other person, but you recognize this is not productive or healthy for me, so I'm going to step away\dots It took me a long time to realize that harm none also means don't put myself in harm's way.'' }
\end{displayquote} Emphasizing the importance of boundary setting in the co-design, P1 put it succinctly: \textit{``Empathy can be exhausting.''}

\subsubsection{Design Concepts to Support Intentionality}

\begin{figure}
    \centering
    \includegraphics[width=.2\textwidth]{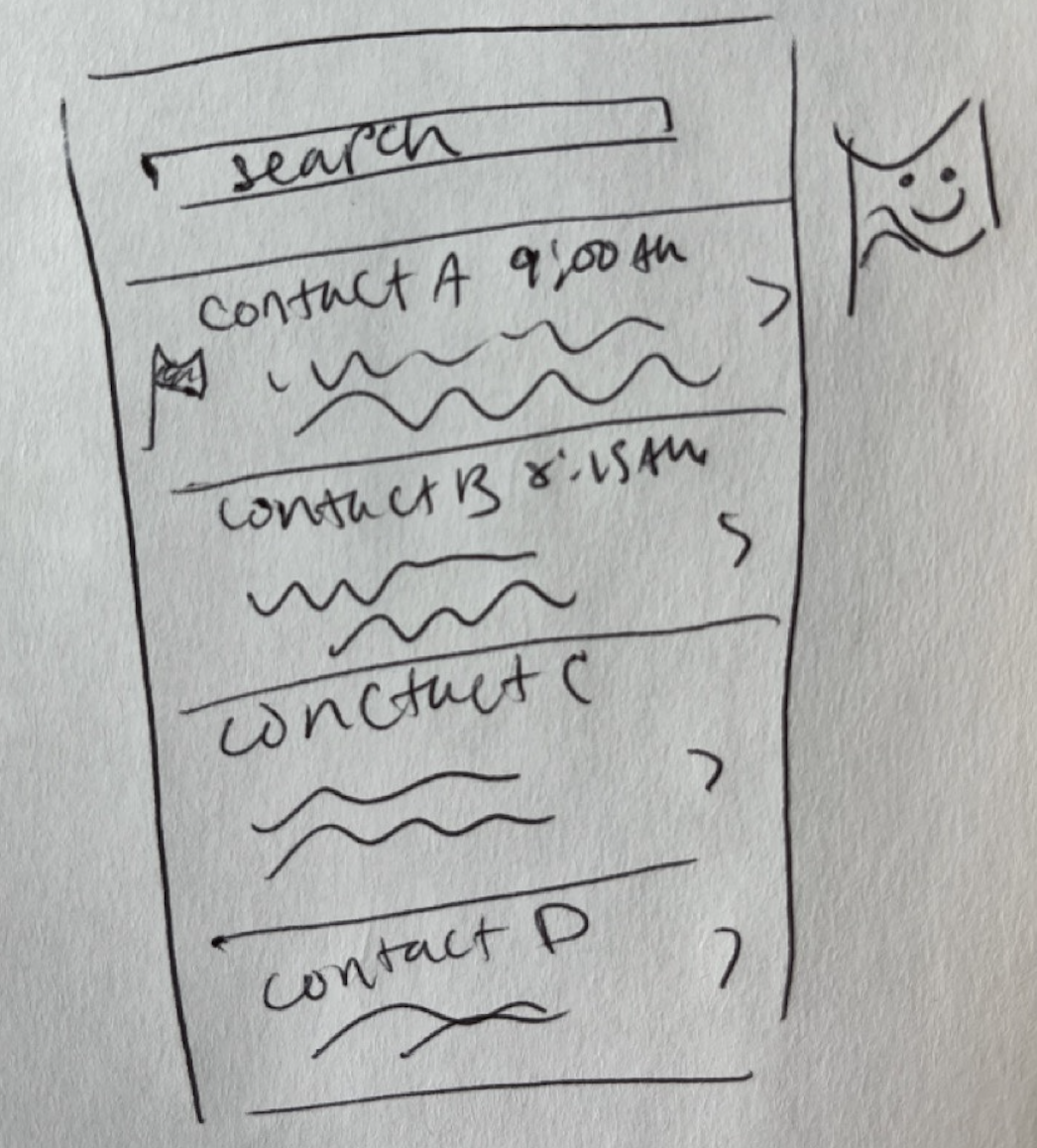}
    \caption{P12 envisioned an ``Empathy Flag'' to signal when they needed empathy from another person over email.}
    \label{fig:empathy_flag}
\end{figure}

Diary study participants envisioned designs that would increase their intentionality by decreasing the speed of their response, and conversely designs that would decrease their intentionality by increasing the speed of their response. As shown in Figure \ref{fig:intentionality_axis} (top right), P10 envisioned slowing themselves down with calming visual elements: \textit{``Some sort of calming image that pops up would be great. You know, like an ice cube\dots I think it would be fun if it just melted from the top to the bottom, something to kind of interrupt the pattern.''} Conversely, participants believed designs forcing them to respond more quickly would inhibit intentionality. As shown Figure \ref{fig:intentionality_axis} (top left), P12 sketched a design that deleted messages if they did not respond within a time limit, noting \textit{``I like to read things over sometimes and give them more thought when I'm trying to be more empathetic. But I don't necessarily get the amount of time \ldots I don't use Snapchat because of this.''} Participants explained that the value of slow designs was to give the user the time to consider their own needs and feelings. During their co-design, P1 envisioned barriers to force them to assess their emotional state before using social media: 

\begin{displayquote}
    \textit{It'd be like five questions \dots and you can answer them quickly \dots but it's also \dots let's take a pre-K pause\dots if you realize, you know what, I don't feel like doing this, then maybe you realize I don't feel like being on social media. Be kind of neat if there's a little `leave' button.}
\end{displayquote}

\noindent Similarly, P2 preferred designs \textit{``where you either freeform write down your thoughts, or have like guiding prompts to help you reflect \ldots AI could be used in some cases when you just need help with having a sounding board.''} Participants' designs also prioritized being intentional about sharing needs. P12 envisioned an \textit{``Empathy Flag''} (Figure \ref{fig:empathy_flag}) to signal they needed empathy when communicating via email.

\begin{figure*}
\centering
\includegraphics[width=.8\textwidth]{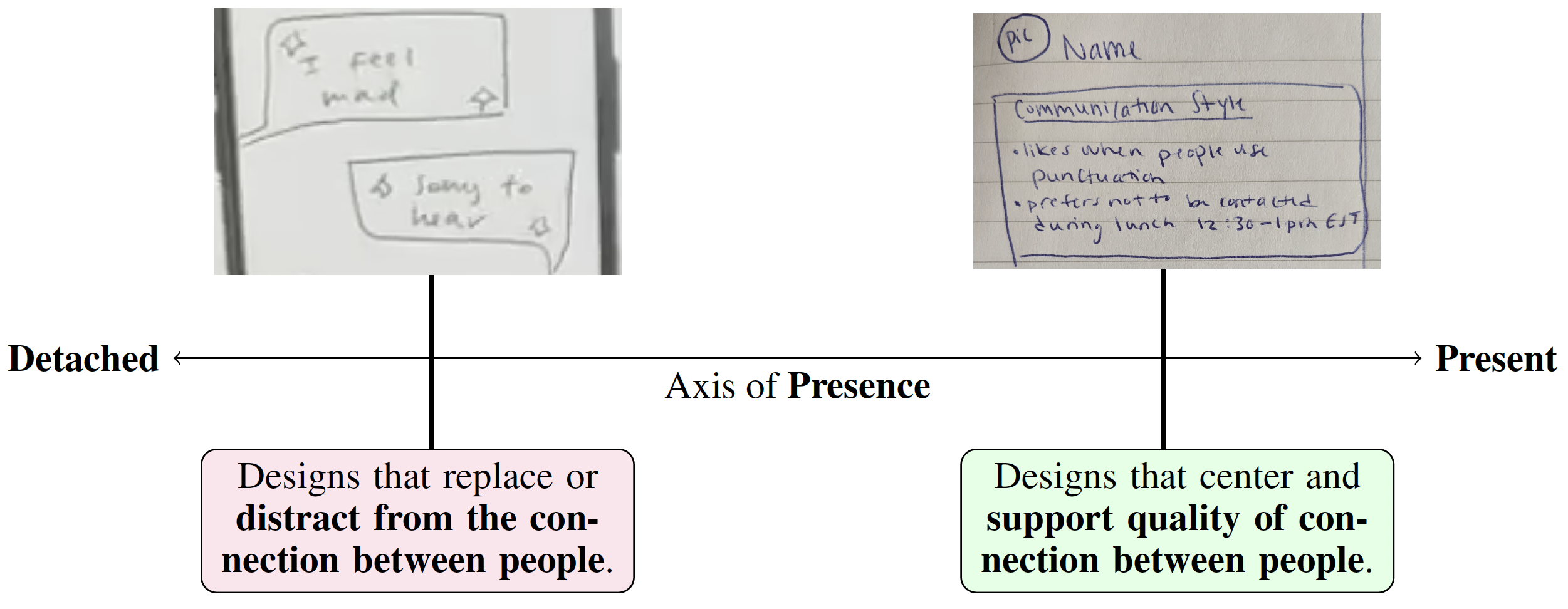}
\caption{Needs-Conscious Design \textbf{supports empathetic presence} achieved between people, and \textbf{eschews design that substitutes human connection with artificial companionship}, or which fragments attention such that one cannot achieve presence with another person. \textbf{Top right}: Co-design drawing with P5, of a cheat sheet of how to improve connection with them based on their communication style. \textbf{Top left}: Co-design drawing with P9, who envisioned a scenario wherein generative AI (responses indicated using stars) entirely removes human presence from the interaction.}
\label{fig:presence}
\end{figure*}

\subsection{Pillar 2: Designing for Presence}

Participants described feelings of shared presence as essential for empathetic connection online, explaining that visible effort fosters sense of presence; that presence can be extended in low-stakes and high-stakes situations; and that spaces for presence must be consciously fostered online.

\subsubsection{Empathetic Presence and Effort Online}

Trainers said that empathetic presence is at the heart of NVC. T14 explained, \textit{``the purpose of NVC is to create a high quality of connection. And once we're connected, it's easier to want to contribute to one another's wellbeing.''} Diary study participants associated empathy online with feelings of another person being present, which often required explicit signals of the attention and effort being offered. In the exit survey, P8 described a friend's Facebook post that elicited empathy, noting that its length and emotional honesty had prompted P8 to reply with an encouraging comment. Conversely, P12 described in a diary entry a time someone did not engage with them: \textit{``Someone was asking for help. I told the person I always got them but they found out I was [at] an event, and the conversation went nowhere\ldots I was unsure what happened so I felt confused \ldots they thought I was too busy to hear them out.''} P10 described the effect of low-effort communication: \textit{``I noticed that Instagram story reactions and direct messages in response to a story can sometimes be hard to respond to with empathy, because they feel cheap.''} 

\subsubsection{Low-Stakes and High-Stakes Presence} T12 noted the importance of offering presence, even when stakes are low: \textit{``Empathy is really about presence\dots if I'm just present with what's going on inside of you\dots that alone is going to feel good.''} When diary study participants reported receiving empathy, it sometimes occurred unexpectedly in low-stakes situations. In their exit survey, P9 described the time they felt most empathized with as \textit{``the instance yesterday when I was telling my friend about how I ate too much pizza\dots really trivial but stood out because he would usually make fun of me for something like this.''} Participants also noted presence is important for high-stakes emotional situations. P7 described such a situation: \textit{``I was going through quite a hard time after having lost my an uncle\dots I needed at least someone who could talk to me almost all the time\dots I approached a friend of mine who, we also don't communicate so much all the time, but then I may say he showed empathy to me\dots and concentrated on me\dots about one week where we made such constant communication and it kept me feeling better.''}

\subsubsection{Creating Spaces for Presence}

Participants said presence had to be consciously fostered online, and described creating spaces for meeting emotional needs. In a diary entry, P4 noted \textit{``during a conversation with my friend, we were discussing a personal challenge I was facing, and my friend created a safe and non-judgmental space for me to express my emotions and share my needs.''} Many participants remarked on the challenges of physical distance. In the entry interview, P13 said of their family, from whom they were spatially distant, \textit{``I don't stay over there with them, so we try to talk online. We try to communicate in whatever way we can\dots trying to explain our side of the story, what we're actually going through.''} P9 shared a similar experience: \textit{``my boyfriend\dots he's long distance\dots when we message each other I have to express empathy, and he also shows empathy to me \dots [if] it's been a stressful week, then I think we relate empathy through\dots text.''}

\subsubsection{Design Concepts to Support Presence}

\begin{figure}
    \centering
    \includegraphics[width=.35\textwidth]{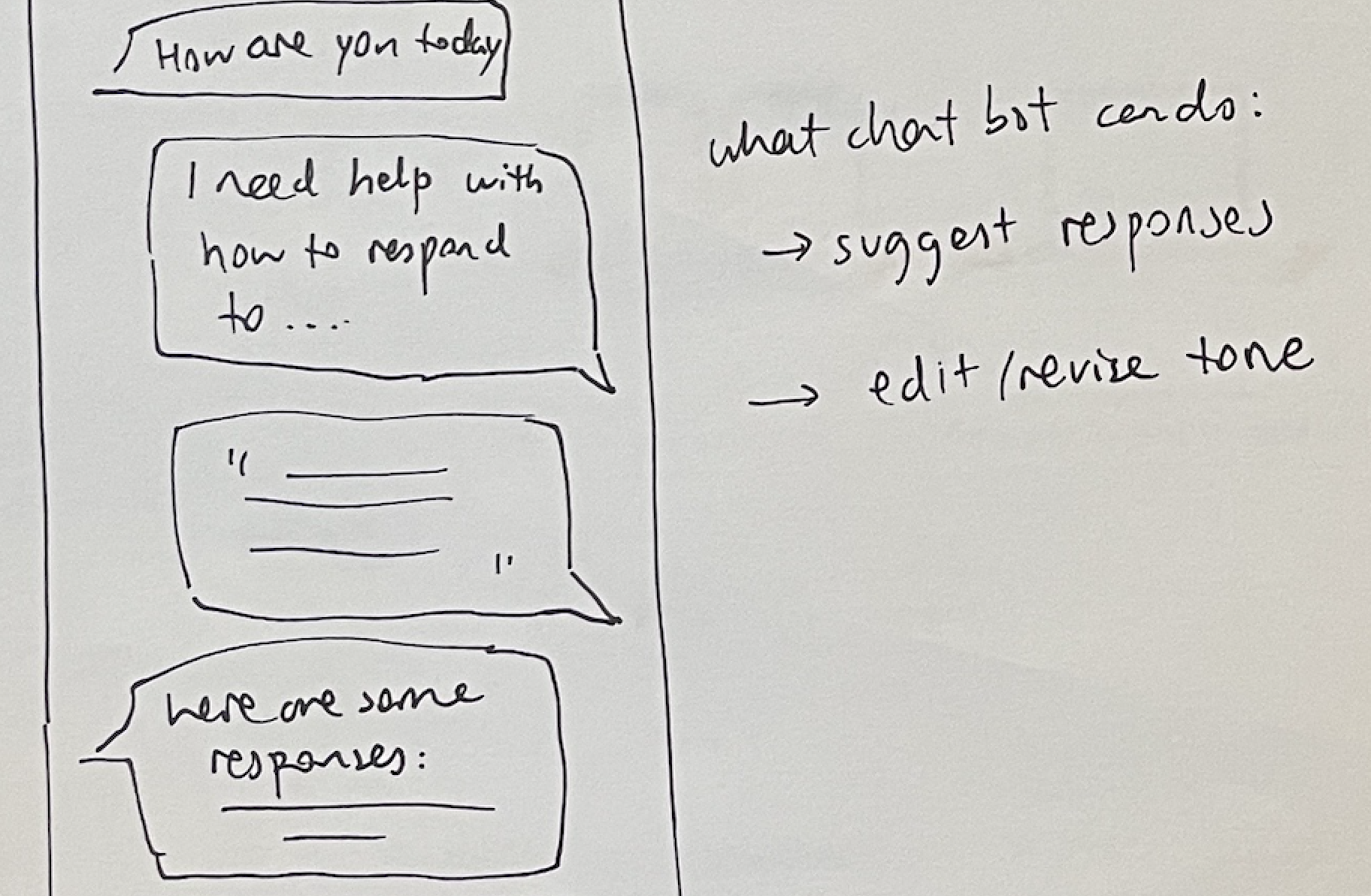}
    \caption{A chat assistant envisioned by P2 to help understand social situations where needs may not be obvious.}
    \label{fig:chat_assist}
\end{figure}

Prioritizes human to human connection proved a chief concern for both trainers and diary study participants. Trainers expressed discomfort that advances in generative AI could undermine human presence online. T4 said, \textit{``as we enter the age of AI, there's a part of me that doubts it's even you\dots It could affect empathy and connection in a potentially dangerous way.''} T7 connected false sense of connection on social media to that offered by generative AI: \textit{``I think about people having connection on Instagram or Facebook, I think when we do that, something inside kind of makes a little checkmark, like I'm connected. I've had intimacy here. But I think the reality of it is really different\dots we're fooling ourselves at the nourishment we're getting. And we don't understand the nourishment we're lacking. And I think the same thing is true about AI.''} T10 expressed optimism about generative AI \textit{``translating things that are more disconnecting into more connecting language''} but noted that \textit{``if it's one AI talking to another AI, it defeats the purpose.''} Despite T4's skepticism about AI-\textit{mediated} communication, they noted that AI could support better connection to oneself: \textit{``even though it's not even a real person on the other end, the bot is helping me to connect to my real experience\dots so a reflection from a bot might support, in some cases, a deeper quality of self-connection.''}

Participants expressed enthusiasm for designs that foster authentic presence. As shown in Figure \ref{fig:presence} (top right), P5 sketched a design to help consciously consider whether another person can offer their presence online: \textit{``here's the important things to know before contacting me\dots instead of letting you just type a message \dots you to have read the first part before you actually\dots start typing.''} The design includes times they are open to communication and styles of communication that would help establish presence. Participants rejected designs that \textit{replaced} human presence. In a design to inhibit empathy (Figure \ref{fig:presence}, top left), P9 sketched an AI system that detects when someone needs empathy, recommends a response, and drafts it. P9 said, \textit{``it would kind of be bad if \dots people only rely on this technology.''} P2 suggested that, instead of composing messages, AI could help interpret difficult social situations, increasing a user's ability to connect. Remarking on the design (Figure \ref{fig:chat_assist}), P2 described an \textit{``AI chatbot inside of Messenger\dots  if you're dealing with a situation that you're not sure how to respond to, you could consult that chatbot\dots copy and paste message context and share it with the chatbot \dots it could give you ideas.''} \looseness=-1

\begin{figure*}
\centering
\includegraphics[width=.8\textwidth]{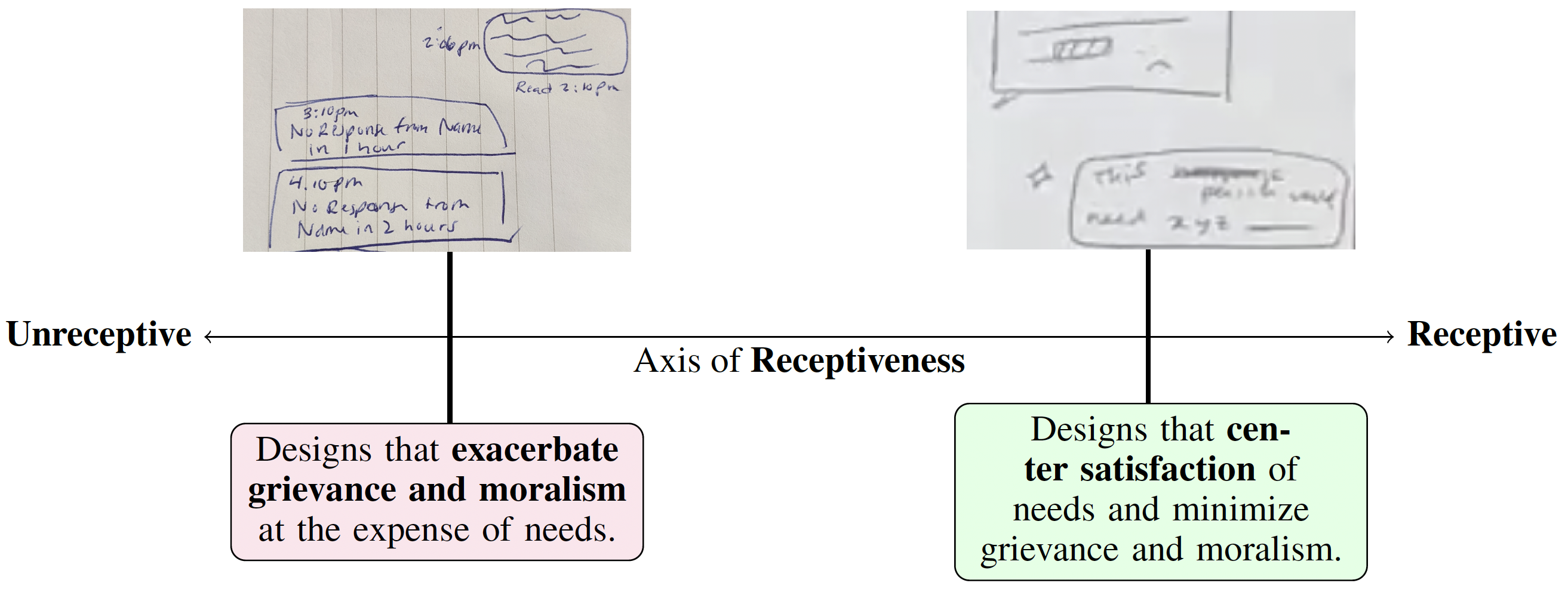}
\caption{Needs-Conscious Design emphasizes  \textbf{receptiveness to needs, both of another person and of oneself}. Being receptive involves nonjudgmental listening and setting aside grievance to hear what someone else needs, and what one needs oneself. \textbf{Top right}: Co-design with P9, who envisioned a generative AI technology that could highlight what another person's underlying needs might be. \textbf{Top left}: Co-design with P5, who envisioned a technology that consistently reminded them that a message they sent had been seen but not responded to, reinforcing the sense that they had been wronged.}
\label{fig:receptiveness}
\end{figure*}

\subsection{Pillar 3: Designing for Receptiveness to Needs}

Trainers and diary study participants noted the importance of clearly expressing needs, and of setting aside notions of who was right and wrong in order to meet needs.

\subsubsection{Centrality of Needs}

CNVC trainers emphasized that needs underlie feelings and serve as the point of connection with others. T1 highlighted the importance of being receptive to needs for achieving connection with others and with oneself: \textit{``People who have maybe never, one, known that they had needs; two, thought that needs were okay to have; and three, ever expressed their needs, begin to acknowledge, and accept, and express [them]\dots that opens up a real connection in relationships.''} T10 describes awareness of needs as a bridge to connecting with others: \textit{``here's this human being, just like me, trying to meet some needs that we share \dots we have the same humanity because we're trying to meet the same needs.''} T14 also noted the importance of \textit{``differentiating between stimulus and cause. The thing that happened outside is a stimulus, but it's not the cause of my feelings.''}

\subsubsection{Clear Expression of Needs}

Trainers and diary study participants highlighted the importance of clearly expressing needs. T14 emphasized a \textit{``mature emotional vocabulary''}, noting \textit{``I can use my words to block you out and not let you know what's going on inside me, or I can use my words to give you a sense of what my experience is.''} T12 encouraged trainees to \textit{``memorize the needs and feelings vocabulary. I think that's really important.''} T7 described learning NVC as analogous to learning a new language, and noted they use a deck of cards \cite{grok2024} containing an \textit{``inventory of feelings and an inventory of needs''} with trainees.

In the entry interview, P1 said \textit{``I see a lot of empathy exhibited in a bisexual group. It's a social community, and there'll be some thirst posts \dots but there's also posts from people who are going through it with their families, and they'll make it clear what they need. Like, hey guys, I really just need to talk to someone about this and probably some of you have gone through this. So I think that that preamble of stating that, hey, I have an emotional need \dots is super helpful.''} P3 noted in the exit survey that online exchanges can facilitate more direct communication of needs, given an unambiguous tone: \textit{``With online communication, I feel like I don't need to beat around the bush \dots as long as there are words that indicate emotions ( `this is so exciting!' , `I love it so much!', `thank you!') then the reader will not view my messages as threats.''} Finally, P8 reaffirmed in their exit survey the importance of a good vocabulary for expressing needs: \textit{``Language and expression is quite important. Sometimes emoticons can be used to convey what words don't.''}

Conversely, participants noted difficulty connecting with others who do not clearly communicate feelings and needs. In the exit survey, P9 said, ``It's harder when people don't respond\dots it's also hard when even if I follow up, they don't clarify any more.'' These issues were magnified when participants received no response at all. P10 described a time that their that request for a Dungeons and Dragons group to meet less regularly went unresponded to: ``\textit{My input wasn't considered\dots I felt irritated that I wrote to a specific person, the game leader, in the group chat, but another individual responded. I felt disappointed that my concerns were ignored.}'' In the exit survey, P3 described ``ghosting'' as a central challenge to empathetic communication online: ``\textit{I wrote a message, then wrote follow up messages, and got no replies. In one case, I got a reply about 4 months after.}''

\subsubsection{Needs vs. Blame}

CNVC trainers stressed focusing not on who was right and wrong but on unmet needs. T4 said, \textit{``we've been conditioned is to focus on who's right and who's wrong\dots and we shift the question\dots rather than who was wrong, who needs what?''} T10 noted NVC \textit{``isn't about making them wrong and bad\dots it's more saying\dots the way you're meeting your needs is not meeting mine.''} T8 describes a strategy \textit{``to translate judgments to needs. The need is the opposite of the judgment most of the time, especially if it's a negative judgment\dots for example, I felt betrayed. Okay\dots I can translate that to \dots I have a need for trust. Trust would be the opposite of the idea of betrayal.''} T2 notes that failing to make this translation, and instead assigning blame, comes at a price: \textit{``what's the cost of blame?\dots blame brings defensiveness and disconnection.''} Diary study participants also noted that feelings of right and wrong inhibit empathy. P11 noted in the entry interview, \textit{``you tend to be less empathetic to others because you have like this baseline of, well, I think this is right and this is wrong.''}

\subsubsection{Design Concepts to Support Receptiveness}

Generative AI informed participant designs for receptiveness to needs. T2 envisioned using generative AI to build awareness of when one's words might fail to meet someone's needs: \textit{``train an AI with that \dots what would it be [like] for you to receive what you're about to say? Just that mirror. Most people are able to say, yeah, I don't think I would like it.''} T7 noted AI might help connect people to their own needs: \textit{``AI is going to be really good at that, not injecting the make-wrong thing, helping people to re-regulate their nervous system. And while they're doing that, during the pause, they're giving themselves empathy. So they're connecting to their needs and their feelings and hopefully doing the same for the other side.''} P9 envisioned AI (Figure \ref{fig:receptiveness}, top right) highlighting needs in a chat, helping them respond empathetically without reducing their effort. Conversely, P5 envisioned AI (Figure \ref{fig:receptiveness}, top left) highlighting that their latest message was left at seen, exacerbating anxiety and sense of grievance. P5 said, \textit{``If you've said something very vulnerable \dots then like, there's just no response\dots you might've completely forgotten about it, and then it just brings it to the forefront.''}
\section{Discussion}

We discuss the three pillars of Needs-Conscious Design, and identify a problematic emergent property of AI-mediated communication identified by our participants. Finally, we provide guiding questions for ensuring \textit{consentful} Needs-Conscious Design, given the potential for interpersonal technologies that can handle user emotional data and impact relationships. 

\begin{figure}[h]
 \centering
\includegraphics[width=.47\textwidth]{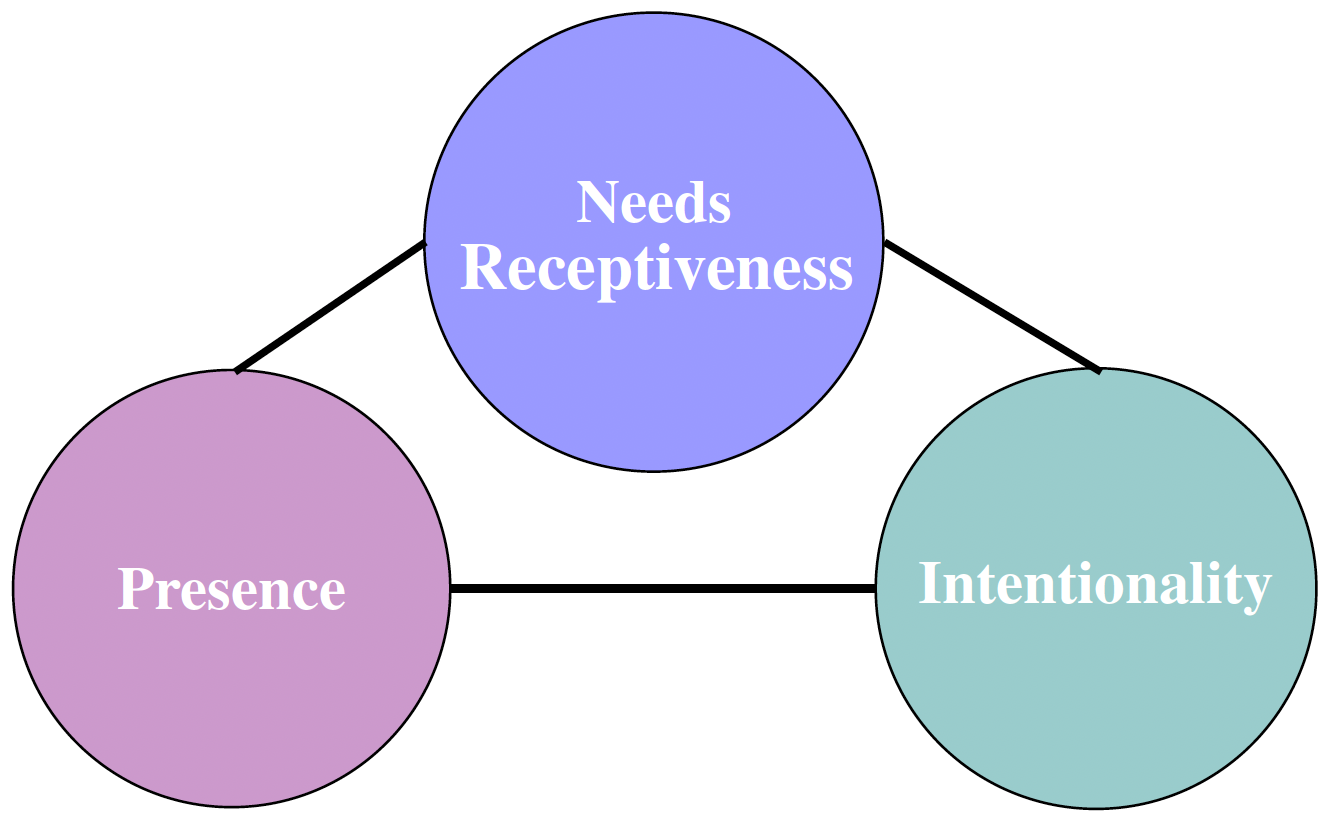}
\caption{Intentionality, Presence, and Receptiveness to Needs form the three pillars of Needs-Conscious Design.}
\label{fig:nvd_vis}
\end{figure}

\subsection{A Conceptual Model of Needs-Conscious Design}

Needs-Conscious Design is supported by the three pillars illustrated in Figure \ref{fig:nvd_vis}: \textbf{Intentionality}, \textbf{Presence}, and \textbf{Needs Receptiveness}. Each pillar depends on the others to facilitate empathetic connection online. Designing for Intentionality ensures that a user feels in control of their words and actions when extending empathy, such that an interaction feels not like the product of the platform but the outcome of the user's carefully considered choices. Designing for Presence ensures that users feel connected to \textit{each other} when they do choose to connect, rather than an artificial agent or other substitute. Designing for Needs Receptiveness ensures that, when users choose to connect, their attention is focused on emotional needs that can be satisfied via empathetic interaction, rather than considerations of right and wrong that obscure unmet needs. Needs-Conscious Design presents three central considerations to the designer, drawing on each of its core pillars:\looseness=-1

\begin{itemize}
    \item \textbf{Intentionality:} Does the design maximize intention and agency, or does it induce responses that might lead a user to feel that they did not have a choice in the matter?
    \item \textbf{Presence:} Does the design facilitate human-to-human presence and connection, or does it try to replace components of human connection with artificial substitutes?
    \item \textbf{Needs Receptiveness:} Does the design focus attention on human needs, or does it exploit human attention and exacerbate feelings of grievance and blame?
\end{itemize}

\noindent Needs-Conscious Design facilitates connection both between people and with oneself. CNVC trainers made clear that not only is it possible to practice NVC with oneself, but connecting with one's own needs precedes the ability to connect with others. Designs guided by the Needs-Conscious Design framework might thus make a user's connection with themselves a central consideration. Even in cases where the relationship with oneself might be the \textit{only} connection enabled by an application, the pillars of Needs-Conscious Design are still well-suited to considering how to enable users to relate to themselves.

Finally, Needs-Conscious Design limits how much generative AI can contribute in the context of human connection. While AI may help prompt humans to invest effort into their relationships, and to direct attention to human needs, it cannot \textit{meet} emotional needs that necessitate connection with another person or connection with oneself. Generative AI cannot invest time, effort, and attention into a relationship, and automated production of language may imitate the \textit{syntax} of empathetic connection, but it cannot replace the \textit{consciousness} of people who choose to share their interior lives.

\begin{figure*}
    \centering
    \includegraphics[width=.8\textwidth]{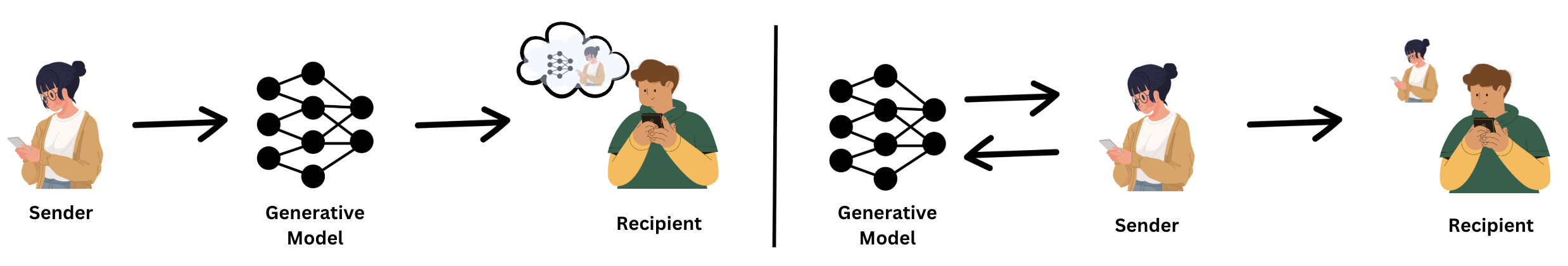}
    \caption{\textit{Empathy Fog} (left), obscuring empathy with AI between sender and recipient; vs. AI for self-reflection (right).}
    \label{fig:empathy_fog}
\end{figure*}

\subsection{\textit{Empathy Fog} in AI-Mediated Communication}

Participants noted that designs using generative AI to \textit{mediate} empathetic connection can \textit{obscure} human attention, inducing uncertainty as to whether the content of a message originated with a human and reflects human effort, or originated with a generative model. This recalls the effortful communication framework of \citet{kelly2017demanding}, which contends that discretionary investment, craft, focused time, responsiveness, and challenge to the sender render communication meaningful. Drawing on this framework, we conceptualize uncertainty induced by AI-mediated communication as \textit{Empathy Fog} (Figure \ref{fig:empathy_fog}, left). Empathy Fog obscures:

\begin{itemize}
    \item Whether the sender has communicated intentionally, or has offered the equivalent of an eloquent auto-reply.
    \item Whether the sender has offered their presence, or used a tool to offer a facsimile of presence, similar to discomfort around AI ``clones'' described by \citet{Lee2023SpeculatingOR}.
    \item Whether the sender is in fact receptive to their needs, and whether the feelings expressed in the sender's message in fact characterize the feelings of the sender.
\end{itemize}

\noindent As noted by \citet{kelly2017demanding}, the effort invested by a sender is meaningful not only to the recipient but to the sender, who feels closer to the recipient for having invested effort. Similarly, Empathy Fog not only obscures the meaning of a message for the recipient, but also for the sender. Without the experience of investing effort in a response, the sender cannot be sure of the presence they have achieved with the recipient. Even when a message accurately reflects the feelings of the sender, the significance conferred by effort for mediating the closeness of the relationship is lost.

On the other hand, engaging in \textit{self-reflection} using the affordances of a generative model (Figure \ref{fig:empathy_fog}, right) can prevent empathy fog and increase a sender's intentionality and receptiveness to needs, commensurate with participant designs to slow down responses and prompt the sender to first consider needs. Rather serving as a mediator between people in interpersonal relationships, the approach that accords most closely with the pillars of Needs-Conscious Design would \textit{mediate a user's relationship with themselves}.

\subsection{Consent in the Context of Needs-Conscious Design}

Generative models may impact user relationships or collect user emotional data \cite{Corvite2023DataSP}. However, Needs-Conscious Design without user \textit{consent} ceases to be nonviolent. Consider the perspective of T7, who said they always asks for consent before practicing NVC with new trainees:

\begin{quote}
    ``\textit{Always ask permission\dots give that person permission, if you get this feeling that I'm doing this thing to you, just please let me know and we can stop right away. So, real permission, giving consent\dots that builds trust\dots honesty and transparency, in my experience with NVC, and other places and other realms, is worth the while. It's worth slowing down for.}''
\end{quote}

\noindent Needs-Conscious Design should yield transparent technologies that enable the user to consent. Even when AI mediates only between a user and themselves, the user should feel they have consented to an interaction that may inform how much of themselves they are willing to offer - including emotional data. We draw on the five dimensions of the affirmative consent framework of \citet{im2021yes} to pose guiding questions for \textit{consentful} Needs-Conscious Design in Table \ref{tab:consent}.

\begin{table}
    \centering
    \small
    \begin{tabular}{|c|p{5.5cm}|}
    \hline
     Dimension    & Guiding Questions \\
     \hline
     Voluntary   & Has the user intentionally adopted the technology? Have they intentionally allowed collection of emotional data (if collected)? \\
     \hline
     Informed   & Does the user know the technology may change their language and affect behavior, even if in a positive way? Does the user know if emotional data is being collected? \\
     \hline
     Revertible   &  If the technology stands alone, can the user remove it? If it is a component of a larger technology, can the user disable it? Can the user permanently erase emotional data?  \\
     \hline
     Specific   & Can the user choose the people with whom they use the technology? Can they select its specific behaviors? Can they specify what forms of emotional data can be collected? \\
     \hline
     Unburdensome & Are the settings easy to access? If a chatbot interface is used, is it trustworthy? Can the user easily manage emotional data? \\
     \hline
    \end{tabular}
    \caption{\footnotesize Questions for \textit{consentful} Needs-Conscious Design.}
    \label{tab:consent}
\end{table}

\subsection{Limitations and Future Work}

We studied Needs-Conscious Design in a broad context, with lay participants who communicated online daily. Future work might study more narrow populations, such as individuals seeking chatbot therapy or romantic relationships with AI companions. We also rely on lay participants for co-designs, where future work might co-design with expert populations such as CNVC trainers. Moreover, we note that NVC is rarely studied in the psychological literature, and has developed into a framework focused on conflict mediation and interpersonal connection, rather than explanatory theories of human behavior. Future work might also interview therapists connect contemporary therapeutic approaches with the methods of NVC. Finally, future work might study diverse cultural contexts, as NVC is practiced around the world \cite{burleson2011assessing}.
\section{Conclusion}

We introduced Needs-Conscious Design, an approach that builds on Nonviolent Communication and centers Intentionality, Presence, and Receptiveness to Needs to facilitate empathy and meet emotional needs. Our research provided design concepts and considerations, while identifying Empathy Fog as problematic property of generative AI. Needs-Conscious Design provides an avenue to supporting human needs, rejecting the substitution of human connection with artificial agents.

\section{Ethical Statement}

Our work surfaces a novel uncertainty in empathetic connection brought on by recent technologies. The findings suggest that, as conversational AI becomes more integrated into ubiquitous communication devices such as cell phones, the effects on human interpersonal connection could be significant. Further research on these topics might help to surface the consequences of technologies that even now are changing the confidence people have in their online exchanges.

\section{Acknowledgments}

This project was supported by the Google Research Scholar program.

\bibliography{references}

\end{document}